# Odd-even staggering of binding energies as a consequence of pairing and mean-field effects

J. Dobaczewski,[1] P. Magierski,[2] W. Nazarewicz,[1,3,4] W. Satuła,[1] and Z. Szymański[1][†]

[1]*Institute of Theoretical Physics, University of Warsaw, ul. Hoża 69, PL-00-681 Warsaw, Poland*
[2]*Institute of Physics, Warsaw University of Technology, ul. Koszykowa 75, PL-00-662 Warsaw, Poland*
[3]*Department of Physics, University of Tennessee, Knoxville, Tennessee 37996*
[4]*Physics Division, Oak Ridge National Laboratory, Oak Ridge, Tennessee 37831*

(November 4, 2018)

Odd-even staggering of binding energies is studied in finite fermion systems with pairing correlations. We discuss contributions of the pairing and mean-field to the staggering, and we construct the binding-energy filters which measure the magnitude of pairing correlations and the effective single-particle spacings in a given system The analysis is based on studying several exactly-solvable many-body Hamiltonians as well as on the analytical formulas that can be applied in the weak and strong pairing limits.

## I. INTRODUCTION

The odd-even staggering (OES) of binding energies, which reflects stronger binding of even-particle-number systems than their odd-particle-number neighbors, has long been known in atomic nuclei [1] and recently has been observed also in metal clusters [2] and in ultrasmall superconducting grains [3]. However, as discussed recently [4], the apparent similarity between the OES in all these finite many-fermion systems is deceptive. Although two basic physical mechanisms are involved: (i) an effect of spontaneous breaking of the spherical symmetry (Jahn-Teller effect [5]) and, (ii) blocking of pair correlations by an unpaired fermion, the origin of the OES is system-specific.

In small metal clusters, OES is believed to be mainly due to the non-spherical shape of the underlying mean field [6–8] and, thus far, neither the empirical evidence nor the theoretical calculations [9] support the presence of superconductive correlations in metal clusters. In superconducting grains, on the other hand, OES is believed to be predominantly due to a blocking effect caused by the presence of an odd electron (see Refs. [10–17]). In the smallest grains, where the average spacing $d$ of the electronic energy levels becomes comparable to the size of the pairing gap [18], $d \sim \Delta$, the OES must invoke both mean-field and pairing effects. In the case of metallic grains, it is, however, difficult (if possible at all) to pin down the detailed structure of the single-electron level spectrum. Most of the theoretical models applied to the problem of superconductivity in grains use an equidistant level spectrum because of its analytical simplicity (see Refs. [10,11,14,15] and references quoted therein). However, more realistic treatments require a non-uniform distribution of single-particle levels as is done, for example, in Ref. [19] (see also Ref. [20]).

A slightly different situation holds in atomic nuclei due to the presence of two types of fermions. The strong and attractive effective proton-neutron interaction gives rise to the appreciable symmetry energy contribution [$\sim(N-Z)^2$] to OES. Because the symmetry energy smoothly varies with particle number and is almost insensitive to shell effects [4], the standard way of extracting the OES has been by means of the higher-order binding-energy indicators (filters), such as the four-point expression of Refs. [21,22] (see also Ref. [24] and references therein). This classical reasoning, which stems from liquid-drop or Fermi-gas models, has recently been questioned [4]. Namely, it has been demonstrated, using fully self-consistent Hartree-Fock theory, that the contribution to OES due to symmetry energy is, in fact, nearly cancelled out by the contribution coming from the smoothed single-particle energy.

The aim of this work is to demonstrate the consistency of our new interpretation using the results of simple, exactly solvable models. Section II discusses the binding-energy indicators used in this study. It also presents the systematics of experimental pairing gaps obtained using these filters. The many-body models employed in this work (seniority model, equidistant-level model, and pairing-plus-quadrupole model) are described in Sec. III, together with the results of the calculations. A short summary and conclusions are given in Sec. IV.

## II. ODD-EVEN STAGGERING FILTERS

The simplest way to quantify the OES of binding energies is to use the following three-point indicator:

$$\Delta^{(3)}(N) \equiv \frac{\pi_N}{2}[B(N-1) + B(N+1) - 2B(N)], \quad (1)$$

---

[†]Deceased September 5, 1999

where $\pi_N = (-1)^N$ is the number parity and $B(N)$ is the (negative) binding energy of a system with $N$ particles. In expression (1), the number of protons $Z$ is fixed, and $N$ denotes the number of neutrons; i.e., this indicator gives the neutron OES. An analogous proton OES indicator is obtained by fixing the neutron number $N$ and replacing $N$ by $Z$ in (1).

By applying filter (1) to experimental nuclear binding energies, $B_{\text{exp}}$, one obtains the experimental neutron and proton OES, $\Delta^{(3)}_{\text{exp}}$. Similarly, by applying this filter to calculated binding energies, one obtains information about theoretical results for the OES. In the following, the same subscripts are used to denote the binding energies and values of indicators (1) obtained for a given model. For example, for the single-particle model described below, the resulting values are $B_{\text{sp}}$ and $\Delta^{(3)}_{\text{sp}}$. Note that by using filter (1) we only aim at facilitating the comparison of calculations with data; however, in essence we always compare and analyze experimental and calculated binding energies.

As a simple exercise, let us first calculate the OES for a system of particles moving independently in a fixed deformed potential well. In such an extreme single-particle model, the binding energy is

$$B_{\text{sp}}(N) = 2 \sum_{k=1}^{\Omega} N_k e_k, \tag{2}$$

where $N_k$ (=0, 0.5, or 1) and $e_k$ stand, respectively, for the occupation number and single-particle energy of the $k$-th, two-fold degenerate level, and the particle number $N$ is given by

$$N = 2 \sum_{k=1}^{\Omega} N_k. \tag{3}$$

In the above expressions, the single-particle energies $e_k$ ($k=1,2,...,\Omega$) appear in ascending order ($e_1 \leq e_2, \ldots, \leq e_\Omega$), and $\Omega$ denotes the number of twofold degenerate single-particle levels.

For the ground state of an even-$N$ system the $N/2$ lowest levels are filled. In the neighboring odd-$N$ system, the odd particle occupies the lowest available level. This implies:

$$\Delta^{(3)}_{\text{sp}}(N = 2n+1) = 0 \tag{4a}$$

$$\Delta^{(3)}_{\text{sp}}(N = 2n) = \tfrac{1}{2}(e_{n+1} - e_n). \tag{4b}$$

Hence, in the absence of the two-body interaction, filter (1) vanishes at odd particle numbers, and it gives half of the single-particle level spacings at even particle numbers. Therefore, following Ref. [4], we assume that any other effect leading to the OES *beyond* the pure single-particle (mean-field) contribution is characterized by values of (1) at odd particle numbers. We also assume that such an effect varies smoothly with the particle number, and, therefore, by subtracting filter (1) at odd and even particle numbers, one may obtain information about the single-particle level spacing at the Fermi level. This hypothesis will be verified in Sec. III using several theoretical models.

Consequently, at odd particle numbers $N = 2n+1$, we shall use the following filter,

$$\Delta(N) \equiv \Delta^{(3)}(N), \tag{5}$$

while at even particle numbers $N = 2n$ we define,

$$\delta e^{\pm}(N) \equiv 2\Delta^{(3)}(N) - 2\Delta^{(3)}(N \pm 1), \tag{6a}$$
$$\delta e(N) \equiv 2\Delta^{(3)}(N) - \Delta^{(3)}(N-1) - \Delta^{(3)}(N+1). \tag{6b}$$

Filters $\delta e^\pm$ correspond to subtracting values of filter (1) at particle numbers higher or lower by one, while $\delta e$ employs the symmetric average of both. The differences between these three filters reflect, therefore, small variations of the OES effects with particle numbers. Up to these small variations, we associate the values of these filters with the spacings of the single-particle levels,

$$\delta e^{\pm}(N) \simeq \delta e(N) \simeq e_{N/2+1} - e_{N/2}. \tag{7}$$

It is clear that filters $\delta e^\pm$ use masses of four nuclides near the given particle number $N$; i.e., they constitute asymmetric expressions in which either heavier or lighter nuclides dominate. On the other hand, $\delta e$ uses masses of five nuclides symmetrically on both sides of $N$. The advantages of using either of these filters depend, therefore, predominantly on the availability of experimental data in the isotopic or isotonic chains. For instance, in Ref. [4] dealing with light- and medium-mass nuclei, the filter $\delta e^+$ was discussed.

It is instructive to relate the asymmetric energy-spacing filters $\delta e^\pm$ of Eq. (6b) to the particle separation energies $S(N)=B(N-1)-B(N)$, i.e.,

$$\delta e^+(N) = S(N) - S(N+2), \tag{8a}$$
$$\delta e^-(N) = S(N-1) - S(N+1). \tag{8b}$$

One sees that the filter $\delta e^+$ depends on separation energies of particles from even systems, while the filter $\delta e^-$ depends on those from odd systems. This does not involve any asymmetry in treating even and odd systems, because, obviously, every particle-separation energy depends on one mass of an even system and on one mass of an odd system.

By using the three-mass filter (5), we hope to avoid mixing the contributions to the OES which originate from single-particle structure with those having other roots, e.g., pairing correlations. Moreover, because it involves three masses only, this filter allows obtaining experimental information on the longest isotopic or isotonic chains. Figures 1 and 2 display experimental values of the neutron $\Delta^{\nu}_{\text{exp}}$ and proton $\Delta^{\pi}_{\text{exp}}$ OES (5). (It is to be noted that these results slightly differ from those presented in Ref. [4]. Firstly, experimental masses were taken from an updated mass evaluation [23]. Secondly, only the masses having an experimental uncertainty less than 100 keV have been considered.)

It is seen that, especially for the light- and medium-mass nuclei, there exists a substantial spread of results around the average trend. This suggests that neutron and proton OES effects are not only functions of neutron and proton numbers, respectively, but that a significant cross-talk between both types of nucleons exists. Numerous studies of this isotopic dependence of the OES exist [24–26], usually based on higher-order filters such as the four-point mass formula [21,22,24]:

$$\Delta^{(4)}(N) \equiv \frac{\pi_N}{4} \left[ 3B(N-1) - 3B(N) - B(N-2) \right.$$
$$\left. + B(N+1) \right] = \frac{1}{2}[\Delta^{(3)}(N) + \Delta^{(3)}(N-1)]. \quad (9)$$

Unfortunately, as demonstrated in Ref. [4], the higher-order filters mix the pairing and single-particle contributions to the OES; hence, they are not very useful for the purpose of extracting the pairing component. Since the detailed analysis of the isotopic dependence of the three-mass filter results is not yet available, below we discuss the average trends; i.e., for each $N$ the values of $\Delta^\nu_{\text{exp}}(N)$ are averaged over all isotones for which the data exist, and, similarly, for each $Z$ the values of $\Delta^\pi_{\text{exp}}(Z)$ are averaged over all isotopes.

Figure 3 (and solid lines in Figs. 1 and 2) present such average values, $\Delta_{\text{exp}}$, for the neutron and proton OES. Neutron and proton values of the OES follow a similar pattern. Namely, they systematically decrease with $A$, and they are reduced around shell and subshell closures, as expected. It is also seen that neutron pairing gaps are systematically larger than the proton ones. The shift is mainly due to a mass dependence of the pairing strength and results from the fact that at a given value $N$, data are more available for lighter nuclei than at an identical value of $Z$. Indeed, for $N\sim Z$, data exist mostly at $A(N)<A(Z)$. A weak contribution from the Coulomb energy is also expected to contribute to this shift.

### III. THEORETICAL MODELS

In this section, we investigate several exactly solvable models to see the interplay between the particle-hole and particle-particle channels of interaction. In all cases, the Hamiltonian has the form

$$\hat{H} = \hat{H}_0 + \hat{H}_{\text{pair}}, \quad (10)$$

where $H = \hat{H}_0$ is either the intrinsic (i.e., deformed) single-particle Hamiltonian or the laboratory-system quadrupole-quadrupole Hamiltonian, and $\hat{H}_{\text{pair}}$ is always the monopole-pairing (seniority) Hamiltonian:

$$\hat{H}_{\text{pair}} = -G\hat{P}^\dagger \hat{P}. \quad (11)$$

In Eq. (11) $G$ is the pairing strength parameter,

$$\hat{P}^\dagger = \sum_{k=1}^{\Omega} a_k^\dagger a_{\bar{k}}^\dagger \quad (12)$$

denotes the monopole-pair creation operator, and $\bar{k}$ denotes the time-reversed state.

Properties of the Hamiltonian (10) depend on the ratio

$$\eta = \frac{G}{\kappa}, \quad (13)$$

where $\kappa$ represents the strength of $\hat{H}_0$. For both $\eta \ll 1$ (weak pairing) and $\eta \gg 1$ (strong pairing), one can treat the Hamiltonian (10) perturbatively. However, the situation encountered most often in the nuclear physics context is the intermediate case ($\eta \sim 0.4$) in which pairing correlations are strongly influenced by the nuclear mean field.

#### A. Limiting cases

To facilitate the discussion of results of exactly solvable models, it is instructive to consider weak- and strong-pairing limits of (10). To this end, we introduce the explicit single-particle Hamiltonian

$$\hat{H}_0 = \hat{H}_{\text{sp}} = \sum_{k=1}^{\Omega} e_k \left( a_k^\dagger a_k + a_{\bar{k}}^\dagger a_{\bar{k}} \right) \quad (14)$$

where $e_k$ are the single-particle energies defined in Sec. II.

##### 1. Weak-pairing limit, $\eta \ll 1$

In this limit, the obvious expansion parameter is $\eta$. The unperturbed ground-state energy, $B^{(0)}_{\eta \ll 1}$, of an even-even system with particle number $N=2n$ ($n$ stands for the number of pairs) is that of Eq. (2), i.e.,

$$B^{(0)}_{\eta \ll 1} = 2\sum_{i=1}^{n} e_i. \quad (15)$$

The pairing Hamiltonian scatters the nucleonic pairs from hole states $(i, \bar{i})$ to particle states $(j, \bar{j})$, and the corresponding scattering matrix element is constant and equal to $-G$. The first order of perturbation theory gives the energy correction

$$B^{(1)}_{\eta \ll 1} = -Gn, \quad (16)$$

while the second-order correction is

$$B^{(2)}_{\eta \ll 1} = -\frac{1}{2}G^2 \sum_{j=n+1}^{\Omega} \sum_{i=1}^{n} \frac{1}{e_j - e_i}. \quad (17)$$

In the case of an odd system with $N=2n+1$, the analogous expressions can be obtained by the following simple modifications. First, in the zero order, the $(n+1)$-th level is occupied (blocked) by one particle, and hence the single-particle energy $e_{n+1}$ should be added to $B^0_{\eta \ll 1}$. Moreover, since the pairing Hamiltonian does not couple orbitals occupied by one nucleon (the blocking effect), the orbital containing the odd particle must be excluded from the sum in Eq. (17), and the number of pairs in Eqs. (16) and (17) must become $n=(N-1)/2$.

Adding together the zero-, first-, and second-order contributions to the binding energy, one obtains the corresponding expressions for $\Delta^{(3)}$:

$$\Delta^{(3)}_{\eta \ll 1}(2n+1) = \frac{1}{2}G$$
$$+ \frac{1}{4}G^2 \left[ \sum_{i=1}^{n} \frac{1}{e_{n+1} - e_i} + \sum_{j=n+2}^{\Omega} \frac{1}{e_j - e_{n+1}} \right], \quad (18a)$$

$$\Delta^{(3)}_{\eta \ll 1}(2n) = \frac{e_{n+1} - e_n}{2} + \frac{1}{2}G$$
$$+ \frac{1}{4}G^2 \left[ \sum_{i=1}^{n} \frac{1}{e_{n+1} - e_i} + \sum_{j=n+1}^{\Omega} \frac{1}{e_j - e_n} \right]. \quad (18b)$$

Finally, for the single-particle energy-splitting filters (6a) and (6b), one obtains:

$$\delta e^+_{\eta \ll 1} = e_{n+1} - e_n$$
$$+ \frac{G^2}{2} \left[ \sum_{j=n+1}^{\Omega} \frac{1}{e_j - e_n} - \sum_{j=n+2}^{\Omega} \frac{1}{e_j - e_{n+1}} \right], \quad (19a)$$

$$\delta e^-_{\eta \ll 1} = e_{n+1} - e_n$$
$$+ \frac{G^2}{2} \left[ \sum_{i=1}^{n} \frac{1}{e_{n+1} - e_i} - \sum_{i=1}^{n-1} \frac{1}{e_n - e_i} \right], \quad (19b)$$

$$\delta e_{\eta \ll 1} = e_{n+1} - e_n$$
$$+ \frac{G^2}{4} \left[ \sum_{\substack{k=1 \\ k \neq n}}^{\Omega} \frac{1}{e_k - e_n} - \sum_{\substack{k=1 \\ k \neq n+1}}^{\Omega} \frac{1}{e_k - e_{n+1}} \right]. \quad (19c)$$

It is clear that in the limit of small $G$ (small compared to the typical single-particle energy spacing), the energy-spacing filters correctly extract the single-particle spectrum from the total binding energies.

### 2. Strong-pairing limit, $\eta \gg 1$

In this case, the expansion parameter is $\eta^{-1}$. In the zero order, the wave function of an even system with $N=2n$ is the ground state of $\hat{H}_{\text{pair}}$; i.e., it corresponds to the state with the maximal quasispin $L_{\max}=\Omega/2$ and the third component of quasispin $L_0=(2n-\Omega)/2$. The binding energy is given by the seniority-model expression (see Refs. [27,28] and Sec. III B),

$$B^{(0)}_{\eta \gg 1}(N=2n) = -G \left[ L_{\max}(L_{\max}+1) - L_0(L_0-1) \right]. \quad (20)$$

In the quasispin formalism, the single-particle Hamiltonian is a combination of a scalar and a vector operator (with respect to the quasispin group), and this implies the $\Delta L = 0, \pm 1$ selection rule for its matrix elements. The expectation value of $\hat{H}_{\text{sp}}$ in the lowest $L=L_{\max}$ state is:

$$B^{(1)}_{\eta \gg 1}(N=2n) = \langle L_{\max} L_0 | \hat{H}_{\text{sp}} | L_{\max} L_0 \rangle = 2n\bar{e}, \quad (21)$$

where

$$\bar{e} \equiv \frac{1}{\Omega} \sum_{k=1}^{\Omega} e_k \quad (22)$$

is the *average* single-particle energy. The second-order correction to the energy is given by

$$B^{(2)}_{\eta \gg 1}(N=2n) = - \sum_{L < L_{\max}, \alpha} \frac{|\langle L_{\max} L_0 | \hat{H}_{\text{sp}} | L L_0 \alpha \rangle|^2}{E^*_{L L_0}}. \quad (23)$$

In Eq. (22) $\alpha$ denotes all the remaining quantum numbers other than $L$ and $L_0$, and $E^*_{L L_0}$ is the unperturbed excitation energy of $|L L_0 \alpha\rangle$. Since $\hat{H}_{\text{sp}}$ can only connect the seniority-zero ground state with the seniority-two, $L=L_{\max}-1$ states at energy $E^* = G\Omega$, Eq. (23) can be reduced to a simple form;

$$B^{(2)}_{\eta \gg 1}(N=2n) = -\frac{1}{G\Omega} \langle L_{\max} L_0 | \hat{H}^2_{\text{sp}} | L_{\max} L_0 \rangle$$
$$+ \frac{1}{G\Omega} \langle L_{\max} L_0 | \hat{H}_{\text{sp}} | L_{\max} L_0 \rangle^2. \quad (24)$$

By introducing the variance of single-particle levels,

$$\sigma_e^2 \equiv \frac{1}{\Omega-1} \sum_{k=1}^{\Omega} (e_k - \bar{e})^2, \quad (25)$$

one can derive a simple expression for the second-order correction:

$$B^{(2)}_{\eta \gg 1}(N=2n) = -\frac{1}{G\Omega^2} 4n(\Omega - n)\sigma_e^2. \quad (26)$$

That is, the first- and second-order corrections to the binding energy are given by the first and second moments of the single-particle energy distribution. Note that for the degenerate $j$-shell, the second-order correction is zero, as expected.

For odd particle numbers, $N=2n+1$, one should take $L_{max}=(\Omega-1)/2$ and $L_0=(2n+1-\Omega)/2$ in Eq. (20). Assuming that the odd particle occupies level $n+1$, this level is removed from the sum of Eqs. (22) and (25), i.e., one needs to consider the remaining $\Omega - 1$ levels only:

$$\bar{e}' \equiv \frac{1}{\Omega - 1} \sum_{\substack{k=1 \\ k \neq n+1}}^{\Omega} e_k = \bar{e} + \frac{1}{\Omega - 1}(\bar{e} - e_{n+1}) \quad (27)$$

and

$$\sigma_e'^2 \equiv \frac{1}{\Omega - 2} \sum_{\substack{k=1 \\ k \neq n+1}}^{\Omega} (e_k - \bar{e}')^2$$
$$= \sigma_e^2 + \frac{1}{\Omega - 2}\left[\sigma_e^2 - \frac{\Omega}{\Omega - 1}(\bar{e} - e_{n+1})^2\right]. \quad (28)$$

The resulting corrections to the binding energies can be written as

$$B_{\eta \gg 1}^{(1)}(N = 2n + 1) = 2n\bar{e}' + e_{n+1} \quad (29)$$

and

$$B_{\eta \gg 1}^{(2)}(N = 2n + 1) = -\frac{1}{G(\Omega - 1)^2} 4n(\Omega - n - 1)\sigma_e'^2. \quad (30)$$

The zero-order expressions for $\Delta^{(3)}$ and $\delta e^\pm$ in the strong pairing limit are given in Sec. III B. By adding the zero- and first-order contributions to the binding energy, one obtains the strong-pairing-limit expressions for $\Delta^{(3)}$:

$$\Delta_{\eta \gg 1}^{(3)}(2n+1) = \frac{1}{2}G\Omega + \frac{\Omega - 2n - 1}{\Omega - 1}(e_{n+1} - \bar{e}) \quad (31a)$$

$$\Delta_{\eta \gg 1}^{(3)}(2n) = \frac{1}{2}G\Omega + \frac{1}{2}G$$
$$+ \frac{\Omega - 2n}{\Omega - 1}\left(\frac{e_{n+1} + e_n}{2} - \bar{e}\right) - \frac{e_{n+1} - e_n}{2(\Omega - 1)}. \quad (31b)$$

It can easily be shown that the first-order correction to $\delta e$ vanishes. Consequently, the seniority-model expressions discussed below give a good approximation to the single-particle energy-splitting filters (6a) and (6b).

### B. Degenerate shell: Seniority model

Let us consider the seniority (or pairing quasispin) model [27,28], i.e., the model for $N$ nucleons moving in a $2\Omega$-fold degenerate shell described by the seniority-pairing Hamiltonian (11). For this model the exact solution can be written in terms of quasispin quantum numbers [see Eq. (20)]. Alternatively, the ground-state energy can be expressed in terms of the seniority quantum number $s$ (see, e.g., [28] p. 222):

$$B_s(N) = -\tfrac{1}{4}G(N - s)(2\Omega - s - N + 2),$$
$$\text{where} \quad \begin{cases} s = 0 & \text{for } N = 2n, \\ s = 1 & \text{for } N = 2n+1. \end{cases} \quad (32)$$

The corresponding value of $\Delta^{(3)}$ is given by:

$$\Delta_s^{(3)}(N) = \begin{cases} \tfrac{1}{2}G\Omega + \tfrac{1}{2}G & \text{for } N=2n, \\ \tfrac{1}{2}G\Omega & \text{for } N=2n+1, \end{cases} \quad (33)$$

which implies

$$\Delta_s(N) = \tfrac{1}{2}G\Omega, \quad (34a)$$
$$\delta e_s^\pm(N) = \delta e_s(N) = G. \quad (34b)$$

It is seen that, in the seniority model, filters (5) and (6) give the OES values (34a) and single-particle energy spacings (34b) which are independent of the particle number $N$. Values of the OES (34a) correctly reproduce the exact pairing gap $\Delta' = \tfrac{1}{2}G\Omega$, which is defined as a half of the lowest excitation energy in an even system, and which we denote by a prime to distinguish it from the OES. The meaning of $\delta e_s$ is less obvious. It is because the mean-field (Hartree-Fock, HF) treatment of Hamiltonian (11) yields only one $2\Omega$-fold degenerate single-particle level at energy $-G$, while our interpretation of $\delta e_s$ assumes that only the Kramers degeneracy is present. Nevertheless, one may compare exact values of $\delta e_s$ with those obtained in an approximated way and see whether the approximate ground-state energies reproduce features of the mass spectrum represented by $\Delta_s$ and $\delta e_s$.

In the seniority model, the Hartree-Fock-Bogoliubov (HFB) equations (which in this case are identical with the BCS equations) can be solved analytically. Indeed, the BCS occupation coefficient is given by

$$v^2 = \frac{N - s}{2(\Omega - s)} \quad (35)$$

(see e.g. [28] p. 233), and the ground-state energy is:

$$B_{s+\text{BCS}}(N) = -\tfrac{1}{4}G(N - s)\left(2\Omega - s - N + \frac{N - s}{\Omega - s}\right). \quad (36)$$

The resulting three-mass filter (1) can be written as

$$\Delta_{s+\text{BCS}}^{(3)}(N) = \begin{cases} \tfrac{1}{2}G\Omega \left[1 + \frac{2N\Omega - N^2 - 2\Omega}{2\Omega^2(\Omega - 1)}\right] & \text{for } N=2n, \\ \tfrac{1}{2}G\Omega \left[1 + \frac{2N\Omega - N^2 - 1}{2\Omega^2(\Omega - 1)}\right] - \frac{G}{2} & \text{for } N=2n+1, \end{cases} \quad (37)$$

and hence filters (5) and (6) give

$$\Delta_{s+\text{BCS}}(N) = \tfrac{1}{2}G\Omega \left[1 - \frac{(\Omega - 1)^2 + (\Omega - N)^2}{2\Omega^2(\Omega - 1)}\right], \quad (38a)$$

$$\delta e_{s+\text{BCS}}^\pm(N) = G\left[1 - \frac{\Omega - 1 \pm (\Omega - N)}{\Omega(\Omega - 1)}\right], \quad (38b)$$

$$\delta e_{s+\text{BCS}}(N) = G\left[1 - \frac{1}{\Omega}\right]. \quad (38c)$$

It is seen that in the limit of large $\Omega$ ($\Omega \gg 1$), the BCS approximation (38a) reproduces the leading order $\mathcal{O}(\Omega^1)$ of the exact OES result (34a). The $\mathcal{O}(\Omega^0)$ deviation from the exact result smoothly depends on $N$ and reaches the minimum ($\sim G/4$) at the middle of the shell ($N=\Omega$).

For the energy-spacing filters the deviations behave as $\mathcal{O}(\Omega^{-1})$. While the symmetric filter $\delta e_{\text{s+BCS}}$ does not depend on $N$, both $\delta e_{\text{s+BCS}}^+$ and $\delta e_{\text{s+BCS}}^-$ vary weakly with the particle number. Namely, at the beginning of the shell ($N \approx 0$), $\delta e_{\text{s+BCS}}^- \approx G$ and $\delta e_{\text{s+BCS}}^+ \approx G - 2G/\Omega$, at the middle of the shell $\delta e_{\text{s+BCS}}^\pm = \delta e_{\text{s+BCS}}$, and at the top of the shell ($N \approx 2\Omega$) $\delta e_{\text{s+BCS}}^+ \approx G$ and $\delta e_{\text{s+BCS}}^- \approx G - 2G/\Omega$. This behavior follows from a simple identity $\delta e_{\text{s+BCS}}^+(N) = \delta e_{\text{s+BCS}}^-(2\Omega - N)$ reflecting the particle-hole symmetry of the model.

The analysis presented in this section illustrates the advantages of comparing exact and approximate results (similarly as experimental and theoretical results) by looking at appropriate filters. Analytical results available in this model allow the explicit study of pairing and mean-field effects. The high degeneracy of the seniority model does not allow for extracting the energy spacings between the deformed levels; to this end, results for the non-degenerate models are shown in the following sections.

### C. Equidistant-level model (infinitely many levels)

Let us consider a phase space of infinitely many, doubly-degenerate, equidistant single-particle levels spaced by $d$. Suppose that all the levels up to a certain Fermi energy are occupied by the fermions, and that they interact through a two-body interaction $\hat{V}$. The Hamiltonian of the model reads,

$$\hat{H} = 2 \sum_{k=-\infty}^{\infty} kd\hat{N}_k + \hat{V}, \qquad (39)$$

where $\hat{N}_k$ is the number operator of the $n$-th level. Furthermore we assume that $\hat{V}$ is invariant with respect to the shift in single-particle indices:

$$V_{k_1 k_2; k_1' k_2'} = V_{k_1+k, k_2+k; k_1'+k, k_2'+k}, \qquad (40)$$

where $k$ is an integer number. For a moment we do not need to specify this interaction; let us only remark that the standard seniority-pairing interaction (11) obviously obeys the shift-symmetry condition.

Of course, we do not intend to consider here infinite numbers of particles and infinite energies, cf. Eq. (39). In practice, we should assume that the number of fermions interacting through $\hat{V}$ is finite but large. The remaining ones (e.g., occupying the most bound shells) form an inert core. This guarantees the effect of the finite spectrum not to be important. Consequently, in future discussions we assume that $N \gg 1$ and that the first level belonging to the "interaction-active" space has $k=1$.

Irrespective of how complicated interaction $\hat{V}$ is, its assumed shift-symmetry allows for an exact analysis in terms of filters based on ground-state energy differences. Indeed, for any particle number $N$, the ground-state energy reads

$$B_{\text{ieq}}(N) = 2 \sum_{k=-\infty}^{\infty} kdN_k(N) + V(N), \qquad (41)$$

where $N_k(N)$ and $V(N)$ are the average values of particle number operators $\hat{N}_k$ and interaction $\hat{V}$ in the ground-state wave functions for $N$ particles. Since the particles interact, values of $N_k(N)$ can be arbitrary numbers between 0 and 1; however, if we switch the interaction off ($\hat{V} \equiv 0$), we revert to the single-particle model (2) in which $N_k = 0$, 0.5, or 1.

From the shift-symmetry (40), it is obvious that the change in the ground-state energy, occurring when the interaction is switched on, is the same for all even systems ($C_{\text{even}}$) and the same for all odd systems ($C_{\text{odd}}$) and hence

$$B_{\text{ieq}}(N) = \begin{cases} B_{\text{sp}}(N) + C_{\text{even}} & \text{for } N=2n, \\ B_{\text{sp}}(N) + C_{\text{odd}} & \text{for } N=2n+1, \end{cases} \qquad (42)$$

where $B_{\text{sp}}(N)$ is given by Eq. (2).

Filters (1), (5), and (6) now give

$$\Delta_{\text{ieq}}^{(3)}(N) = \Delta_{\text{sp}}^{(3)}(N) + C_{\text{odd}} - C_{\text{even}}, \qquad (43)$$

hence

$$\Delta_{\text{ieq}}(N) = C_{\text{odd}} - C_{\text{even}}, \qquad (44a)$$
$$\delta e_{\text{ieq}}^\pm(N) = \delta e_{\text{ieq}}(N) = d. \qquad (44b)$$

It is quite remarkable that the above argumentation does not at all depend on details of the two-body interaction, and that – irrespective of the interaction – filters (5) and (6) correctly separate the interaction effects (44a) from the single-particle spacings (44b).

This generic result does not depend on whether any approximations are used to obtain the ground-state energies of interacting systems. In particular, the BCS mean-field results (obtained for interaction $\hat{V}$) will also obey the pattern presented in Eqs. (44) exactly; only the value of the interaction-energy difference $C_{\text{odd}} - C_{\text{even}}$ may be different from the exact result.

Similarly, the BCS results obtained for the seniority-pairing interaction (11) also follow the same pattern; however, in this case we have to additionally ensure that the phase space in which the pairing correlations are allowed to develop is the same for all particle numbers. Indeed, one usually solves the BCS equations in a finite phase-space window, adjusting the interaction strength $G$ to that window. For the results (44) to be valid, we have to always use the same number of levels below and above the Fermi level, independently of the number of particles. Actually, such a prescription is often used in realistic BCS or HFB nuclear structure calculations.

As discussed above, in the case of the equidistant-level model, the energy-filter $\delta e$ reproduces the value of the level spacing $d$ regardless of the detailed structure of $\hat{V}$ and details of the many-body approximation used. The reason for this is the shift-symmetry of the Hamiltonian. Of course, if this symmetry is broken either by assuming the finite Hilbert space (see the following section) or by introducing the explicit symmetry-violating terms, one cannot *a priori* expect the above conclusion to hold *exactly*.

It is also worth noting that the OES (44a) is defined in terms of the pairing (interaction) energy in the even and odd system, and not in terms of the HFB or BCS pairing gap $\Delta'$ (defined as half of the lowest excitation energy). Only for strong pairing correlations are the pairing gaps and the values of the OES the same (see Ref. [11]). In nuclei, pairing correlations never reach such a limit, and the OES and BCS pairing gap in an odd system can be significantly different.

### D. Deformed shell (finite non-degenerate spectrum)

In this section, we investigate the deformed-shell-plus-pairing Hamiltonian that contains the single-particle term (14) (for $\Omega$ twofold degenerate single-particle states available for the pair scattering) and seniority-pairing interaction (11). In this case, the analytic solution does not exist, but the exact eigenstates can be found numerically using, for example, the Richardson method [29] or by performing a direct diagonalization. The latter approach cannot be applied when the number of single-particle levels is large, because the dimension of the Hilbert space grows as $\binom{\Omega}{\Omega/2}$, and this puts the practical limit at $\Omega \sim 20$. imaginary-time method

A particular variant of this model is when the single-particle spectrum is uniformly spaced (finite-number equidistant-level model), i.e., $e_k = kd$ ($k=1,...,\Omega$). In this work, we are mainly interested in the ability of the energy-spacing filter (6) to extract the single-particle spectrum from the total binding energies. Therefore, below, we also study the case of a nearly-equidistant spectrum, in which $e_k = kd$ except for the seventh level shifted up in energy by $d/4$ (i.e, $e_7 = 7.25d$). The value of the single-particle spacing $d$ constitutes a convenient energy scale, and below, the results will be expressed as ratios of all energies and parameters with respect to $d$.

Figures 4 and 5 show the results of the exact calculations for the nearly-equidistant spectrum of $\Omega=16$ levels. The behavior of the three-mass filter (1) is illustrated in Fig. 4, where values of $\Delta_{\text{def}}^{(3)}$ are compared to

$$\Delta^0 \equiv G\sqrt{\langle \hat{P}^\dagger \hat{P}\rangle - [N/2]}, \qquad (45)$$

which is the "equivalent" pairing gap. (In the BCS limit, $\Delta^0$ becomes the gap parameter.)

For $G/d=0.1$ (weak pairing), values of $\Delta_{\text{def}}^{(3)}(2n+1)$ nicely follow the low values of $\Delta^0$, while those of $\Delta_{\text{def}}^{(3)}(2n)$ are clearly influenced by the single-particle spectrum and do not at all reflect the smallness of the pairing correlations. In particular, the fluctuation around $N=14$ ($n=7$) clearly shows up in $\Delta_{\text{def}}^{(3)}(2n)$ and is absent in $\Delta_{\text{def}}^{(3)}(2n+1)$.

In the case of intermediate pairing ($G/d=0.3$), $\Delta_{\text{def}}^{(3)}(2n+1)$ behaves rather smoothly, while $\Delta_{\text{def}}^{(3)}(2n)$ zigzags in the region of irregularity in the spectrum. No direct correspondence between the values of $\Delta_{\text{def}}^{(3)}$ and $\Delta^0$ can be found here. However, with increasing values of $G$, i.e., when the static pairing sets in, $\Delta_{\text{def}}^{(3)}$ closely approaches $\Delta^0$. This is nicely illustrated for $G/d=0.5$. Only in the case of relatively strong pairing correlations (but still far from the strong pairing limit discussed in Sec. III A 2) is the fluctuation in $\Delta_{\text{def}}^{(3)}$ barely visible.

In the case of weak pairing correlations, the corresponding energy-spacing filters (6), shown in Fig. 5, very well reflect the structure of $\hat{H}_{\text{sp}}$. With an increasing pairing strength, the pairing Hamiltonian gives rise to a diffused Fermi surface (i.e., it smears out the single-particle occupations). Consequently, the information about the details of the single-particle distribution is then expected to be washed out. This is clearly seen in Fig. 5, where the nearly-equidistant single-particle spacings are marked by dots.

However, it is seen that even in the case of relatively strong pairing correlations, ($G/d=0.4$), the symmetric energy-spacing indicator $\delta e$ gives a qualitative description of the spectrum. Even at $G/d=0.5$, the zigzag in $\delta e$ appears in the right place. As far as the asymmetric filters $\delta e^\pm$ are concerned, their behavior is more strongly influenced by $G$, and the resulting particle number dependence may make them less useful measures of the spectral properties.

In order to assess the quality of the quasi-particle (BCS) approximation for the binding-energy indicators, in the same model we have also carried out the BCS calculations. As seen in Fig. 6, it is only at large values of $G$ that the $\Delta^{(3)}$ filters (1) which are applied to the BCS energies approach the exact results. Especially at intermediate values of pairing strength, where the static pairing vanishes in odd-$N$ nuclei, BCS becomes a rather poor approximation. However, it is clear that values of $\Delta_{\text{def+BCS}}^{(3)}(2n)$ are affected by the single-particle spectrum, while those of $\Delta_{\text{def+BCS}}^{(3)}(2n+1)$ are rather insensitive to it.

The agreement is significantly better for the energy-spacing filter $\delta e$ (Fig. 7). Again, as discussed above, at large values of pairing strength, $\delta e$ is only a qualitative measure of the single-particle splitting.

*1. Second-order perturbation results*

Following Sec. III A, one can derive analytic expressions for the binding energy filters which should be valid in the limiting cases of weak and strong pairing. Particularly simple are the weak-pairing expressions. Namely, Eqs. (18) yield:

$$\Delta_{\text{def}}^{(3)}(2n+1) = \frac{1}{2}G + \frac{G^2}{4d}\left[S(n) + S(\Omega - n - 1)\right], \quad (46a)$$

$$\Delta_{\text{def}}^{(3)}(2n) = \frac{d+G}{2} + \frac{G^2}{4d}\left[S(n) + S(\Omega - n)\right], \quad (46b)$$

where

$$S(n) \equiv \sum_{i=1}^{n} \frac{1}{i}. \quad (47)$$

Using Eq. (19), the single-particle splitting filters are:

$$\delta e^+ = d + \frac{G^2}{2d(\Omega - n)}, \quad (48a)$$

$$\delta e^- = d + \frac{G^2}{2dn}, \quad (48b)$$

$$\delta e = d + \frac{G^2}{2dn(\Omega - n)}. \quad (48c)$$

Note that in this case, the single-particle splitting filters (48) are always greater than the original spacing $d$. In particular, $\delta e^+$ increases with $n$, $\delta e^-$ decreases with $n$, and $\delta e$ has a minimum in the middle of the shell.

In the strong-pairing limit of the equidistant-level model, the zero-order expressions for $\Delta_{\text{def}}^{(3)}$ and $\delta e$ are given by Eqs. (33) and (34). The corresponding higher-order corrections can be easily derived using Eqs. (21)-(30) after noting that

$$\bar{e} = \frac{\Omega + 1}{2}d \quad \text{and} \quad \sigma_e^2 = \frac{\Omega(\Omega + 1)}{12}d^2. \quad (49)$$

Figure 8 shows the accuracy of the second-order expressions for the binding energies in the equidistant-level model with $\Omega=16$ in the weak-pairing (left) and strong-pairing (right) limits. It is seen that, for large numbers of particles, the low-$\eta$ expansion is fairly accurate even for $\eta \sim 0.4$. Also the $\eta^{-1}$ expansion seems to work very well even for large values of $d/G$. different vertical

### E. Pairing-plus-quadrupole model

In the previous sections, we have considered models having either degenerate (Sec. III B) or arbitrarily fixed (Secs. III C and III D) single-particle spectra. However, in real systems, the single-particle energies do depend on the numbers of particles, and the energy-spacing filters (6) extract from masses the single-particle energies which include the particle-number dependence, i.e, at every $N$, the distance to the next available single-particle level is obtained.

In order to account for such effects, in this section we analyze the results of the exact diagonalization of the pairing-plus-quadrupole (PPQ) Hamiltonian,

$$\hat{H} = -G\hat{P}^\dagger \hat{P} - \kappa \hat{Q} \cdot \hat{Q}, \quad (50)$$

in a single-$j$ shell. The PPQ model is well known and has been used many times to test collective properties of fermion systems; here we use the version defined in detail in Ref. [30]. Calculations were performed in the $j=19/2$ shell ($\Omega=10$). The strength of the quadrupole-quadrupole (QQ) interaction $\kappa$ provides here a suitable energy scale, and in the following all the energies and parameters are expressed as ratios with respect to $\kappa$.

Exact ground-state energies for all particle numbers $N=0,1,\ldots,20$ and $G=0$ are plotted in Fig. 9. All the even systems have the ground-state spins of $I_{\text{g.s}}=0$, while in the odd systems, the ground-state spins are $I_{\text{g.s}}=|N-\Omega|/2+\Omega/2$; this corresponds to rotational bands based on oblate sequences of deformed single-particle levels (see the discussion in Ref. [30]).

Even in a large energy scale of Fig. 9, the effect of the twofold Kramers degeneracy and the OES effect are clearly seen. Note that in this model the pure QQ interaction generates a (weak) OES effect. Indeed, in a relatively small phase space, the QQ interaction has a tangible pairing component.

In order to quantify the OES and mean-field effects, Fig. 10 shows values of the three-mass filter (1) obtained for different pairing strengths $G$. Ground-state energies of even and odd systems are used to calculate $\Delta_{\text{PPQ}}^{(3)}$ for each value of $G$. For even systems, the ground-state spins equal 0 for all values of $N$ and $G$, while for odd systems, the values of $I_{\text{g.s}}=19/2$ (for all $N$) replace at large $G$ the values of $I_{\text{g.s}}=|N-\Omega|/2+\Omega/2$, which characterize the $G=0$ solutions. This corresponds to a gradual transition from deformed to spherical shapes and from rotational to seniority-like excitation spectra.

In Fig. 10, one can clearly see that the OES increases almost linearly with $G$, while the pattern of alternating larger and smaller values of $\Delta_{\text{PPQ}}^{(3)}$ is almost independent of $G$. Both these features of $\Delta_{\text{PPQ}}^{(3)}$ are explicated by using filters (5) and (6b), which give values of $\Delta_{\text{PPQ}}$ and $\delta e_{\text{PPQ}}$ plotted in Fig. 11. One can very well see the almost linear dependence of the OES on the pairing strength $G$, Fig. 11(a), and a very weak $G$-dependence (apart from $N=10$) of the single-particle energy spacings, Fig. 11(b).

The PPQ model exhibits several features pertaining to two kinds of phase transitions. First, the static pairing correlations set in at critical values of the pairing strength $G$. Depending on the number of particles, this phase transition occurs at about $G/\kappa=0.03$–0.05 for even particle number. Second, the transition from deformed to spherical shapes occurs at slightly higher values of $G$, i.e., at about $G/\kappa$=0.08, 0.10, 0.12, and 0.12 for $N$=4, 6, 8, and 10, respectively. In addition, deformed equilibrium shapes are oblate for $N$=2, 4, 6, and 8, and prolate for $N$=10. (Note that apart from a linear dependence of energies on $N$, the PPQ Hamiltonian is exactly symmetric with respect to the particle-hole transformation $N \leftrightarrow 2\Omega - N$ [31].)

Since the system is finite (and fairly small for that matter), the phase transitions are hardly visible in the exact results of Fig. 11. However, when filters (5) and (6b) are applied to the mean-field (HFB) ground-state energies (Fig. 12), the phase transitions become visible as sudden increases in $\Delta_{\text{PPQ+HFB}}$ (pairing transition), and degeneracies of $\delta e_{\text{PPQ+HFB}}$ (shape transition).

As one can see, a comparison of the exact and HFB ground-state energies (Figs. 11 and 12) is very instructive when it is based on comparing the corresponding filters (5) and (6b). It turns out that in the PPQ model the HFB method reproduces quite well the OES and the single-particle properties simultaneously. Some deviations occur only near the phase transitions, where it is well known that the mean-field approximation is not accurate.

When analyzing exact solutions for systems interacting with two-body interactions, or when similarly analyzing the experimental data, one does not have *a priori* access to the single-particle energies or to the single-particle energy-spacings. In fact, the single-particle energies are concepts that appear naturally only in the mean-field approximation. Therefore, in order to assess the meaning of numbers obtained from the energy-spacing filter (6b), one should compare them with the energy-spacings calculated directly from the mean-field spectra, i.e., with the differences

$$\delta \varepsilon(N) = \varepsilon_{N/2+1}(N) - \varepsilon_{N/2}(N), \qquad (51)$$

[cf. Eq. (7)] where $\varepsilon_k(N)$ is the $k$th (twofold degenerate) single-particle energy obtained within the mean-field approximation. Since a consistent application of the mean-field approximation to an odd system requires the time-reversal breaking, the Kramers degeneracy is lifted in odd systems. Consequently, in Eq. (51) one should only use the single-particle energies obtained self-consistently for systems with an even number of particles.

In Fig. 13(b) we show the differences (51) calculated from the single-particle spectra of canonical HFB energies obtained in the PPQ model. (In fact, since all of the PPQ+HFB equilibrium solutions conserve the axial symmetry, the PPQ+HFB method reduces to the simple BCS approximation, and the canonical energies are equal to the eigenenergies of the mean-field Hamiltonian.) It is clear that the energy-spacing filters (applied either to the exact or to the HFB total energies) give results similar to the differences of canonical energies only for deformed shapes. Whenever the mean-field solutions become spherical, differences (51) collapse to zero, as expected, while the energy-spacing filters give non-zero values. This result is easy to understand: in the spherical limit, the results of the PPQ model should resemble those of the seniority model. While the single-particle energies are degenerate (the Hilbert space consists of one $j$-shell only), the energy-spacing filter should be proportional to the pairing strength $G$ (see Sec. III B). On the other hand, before the transition to sphericity, the results presented in Figs. 11(b), 12(b), and 13(b) are encouragingly similar. Note that deviations obtained at $N$=10 should again be attributed to the differences in predicted shapes (prolate vs. oblate).

Finally, in Fig. 13(a) we show the values of the HFB "equivalent" gap parameters (45) calculated in the PPQ+HFB model. It is seen that $\Delta^0$ significantly underestimates the magnitude of the OES effect, and moreover, it exhibits some particle-number dependence which is absent in the exact results.

In light of the above discussion, the very weak average dependence of $\Delta_{\text{PPQ}}^{(3)}$ on $N$ (except for the OES, of course), shown in Fig. 10, can be given by a very simple explanation. In the weak pairing limit, $\Delta_{\text{PPQ}}^{(3)}$ is small, and its overall particle-number dependence is much weaker than the even-odd effect (see Fig. 6). On the other hand, in the limit of strong pairing, $\Delta_{\text{PPQ}}^{(3)}$ is expected to approach the seniority limit in which $\Delta^{(3)}$ depends only on the number parity but not on $N$. Note that the results of the non-degenerate model shown in Fig. 6 are very far from the spherical limit since $G/d < 1$.

## IV. CONCLUSIONS

This work contains the analysis of an interplay between pairing and mean-field effects on binding energies of many-fermion systems. While most of our discussion is concerned with nuclear systems, the main conclusions also apply to other finite-size superconductors such as grains.

The analysis of binding energies of several exactly-solvable Hamiltonians (allowing variations in the magnitude of pairing correlations) demonstrates that the three-mass indicator, $\Delta^{(3)}(2n+1)$, is indeed an excellent measure of pairing correlations, and the symmetric filter $\delta e$ adequately extracts the effective single-particle spacings from the measured binding energies of *deformed* nuclei. According to the analysis of nuclear masses, the mean-field contribution to the OES is significant in light nuclei, but it is reduced to ~100-200 keV in heavy deformed nuclei [32] due to the relatively close spacing of single-particle levels.

For deformed nuclei with weak and intermediate pairing correlations, there is a nice consistency between the approximate results obtained within the mean-field approach (BCS or HFB) and the exact results. However, for strongly paired nearly-spherical nuclei, there is no clear correlation between $\Delta^{(3)}(2n+1)$ and the pairing deformation $\Delta^0$. Also, the energy-spacing filters are superior over the mean-field single-particle energies in assessing the single-particle properties of the system.

Approximate expressions of binding-energy indicators have been derived in the limits of weak and strong pairing. These formulae nicely explain the gross particle number dependence seen in the exact results.

## ACKNOWLEDGMENTS

This research was supported in part by the U.S. Department of Energy under Contract Nos. DE-FG02-96ER40963 (University of Tennessee), DE-FG05-87ER40361 (Joint Institute for Heavy Ion Research), DE-AC05-96OR22464 with Lockheed Martin Energy Research Corp. (Oak Ridge National Laboratory), and by the Polish Committee for Scientific Research (KBN) under Contract No. 2 P03B 040 14.

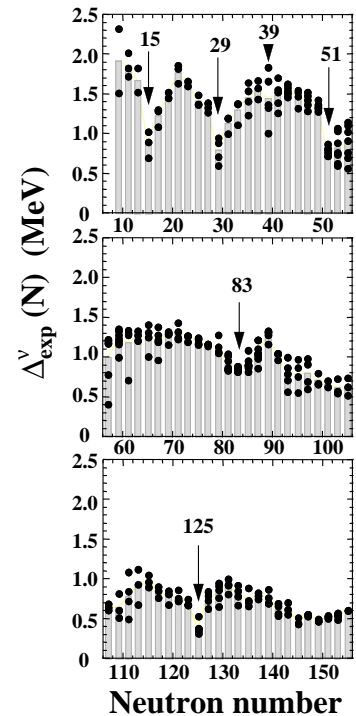

FIG. 1. Neutron OES filter $\Delta^\nu_{\mathrm{exp}}$ (5) plotted as a function of neutron number $N$ ($N$-odd). Results for different isotones are marked by dots. The average values of $\Delta^\nu_{\mathrm{exp}}$ are indicated by gray bars. Experimental data were taken from Ref. [23].

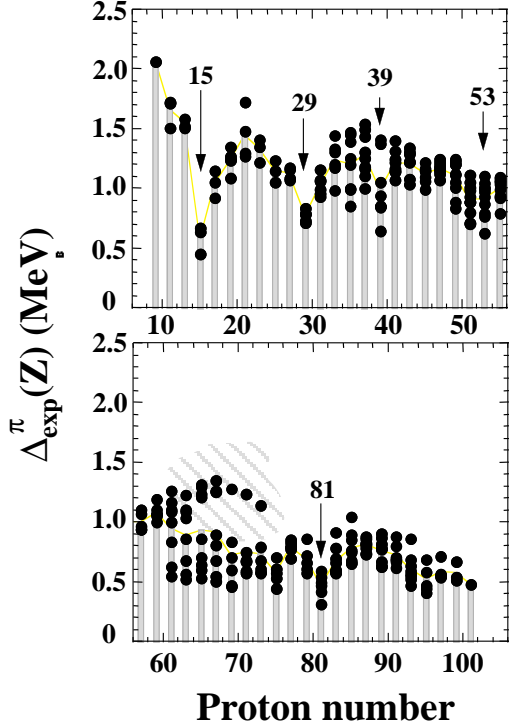

FIG. 2. Same as in Fig. 1 except for the proton OES filter $\Delta^\pi_{\exp}$ (5) plotted as a function of $Z$ ($Z$-odd).

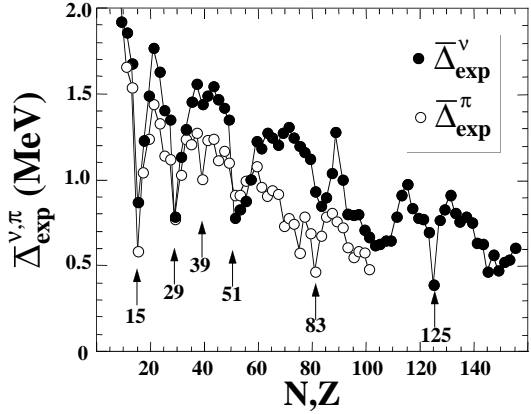

FIG. 3. Average values of neutron and proton OES filters (5) plotted as functions of the neutron and proton numbers (both $N$ and $Z$ odd), respectively. Experimental masses were taken from Ref. [23].

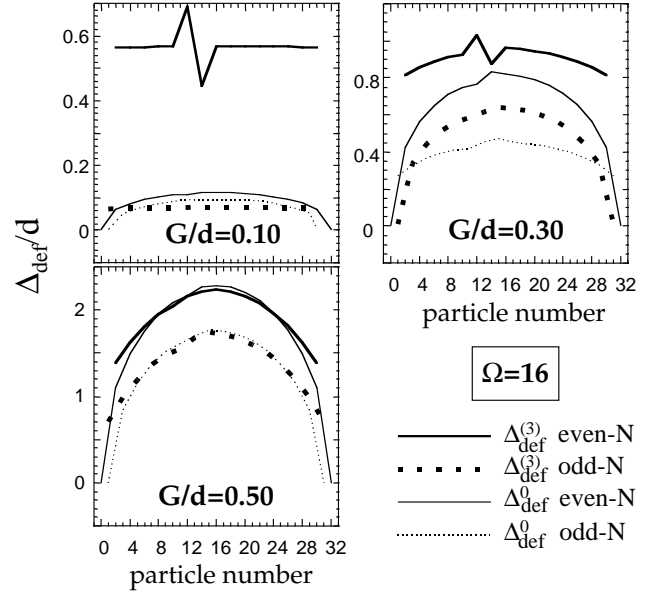

FIG. 4. Three-mass filters $\Delta^{(3)}_{\mathrm{def}}$ (1) (thick solid and dotted lines) calculated for the exact binding energies in the deformed-shell-plus-pairing model with $\Omega=16$ for the case of weak ($G/d=0.1$), intermediate ($G/d=0.3$), and strong pairing ($G/d=0.5$). The single-particle spectrum is uniform ($e_k = dk$), except for the seventh level which is shifted up in energy by $d/4$ (i.e, $e_7 = 7.25d$). The equivalent gap parameters (45) are shown by thin lines.

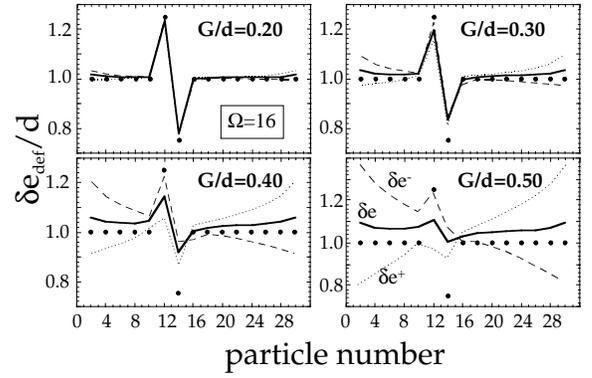

FIG. 5. Similar to Fig. 4 except for the energy-spacing filters (6) $\delta e_{\mathrm{def}}$ calculated for $G/d=0.1$, 0.2, 0.3, and 0.4. The nearly-equidistant single-particle spacings are marked by dots.

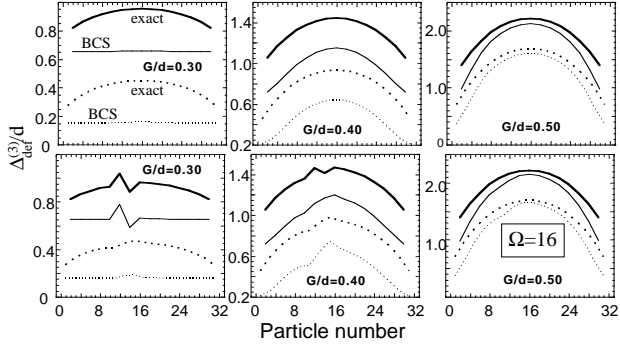

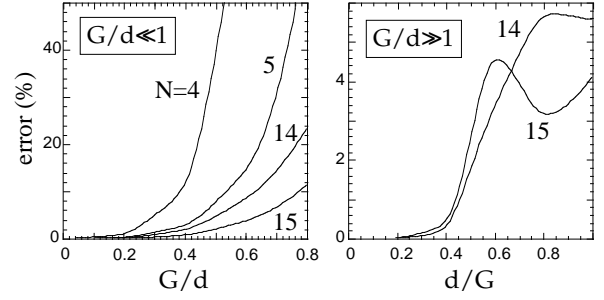

FIG. 6. Three-mass filters $\Delta^{(3)}_{\rm def}$ (1) calculated for the exact (thick lines) and BCS (thin lines) binding energies in the deformed-shell-plus-pairing model with $\Omega=16$ and a nearly-equidistant spectrum. Solid and dotted lines show results for even and odd values of $N$, respectively.

FIG. 8. Relative error (in percent) of the second-order expressions for the binding energies of the equidistant level model with $\Omega=16$ in the weak (left, $G/d\ll 1$; Sec. III A 1) and strong (right, $d/G\ll 1$; Sec. III A 2) pairing limits. In the weak pairing limit, calculations were performed for $N=4$, 5, 14, and 15. To realize the strong-pairing situation, only large particle numbers, $N=14$ and 15, were considered in the $d/G\ll 1$ case.

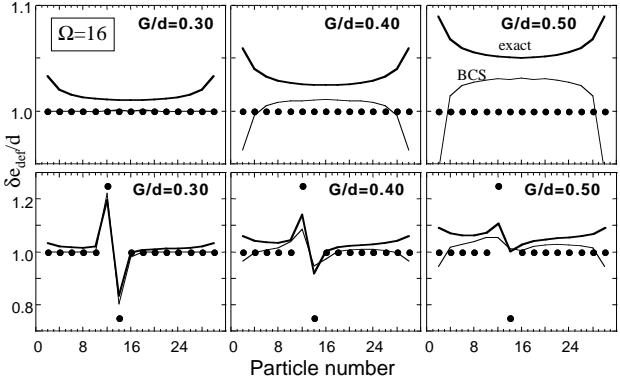

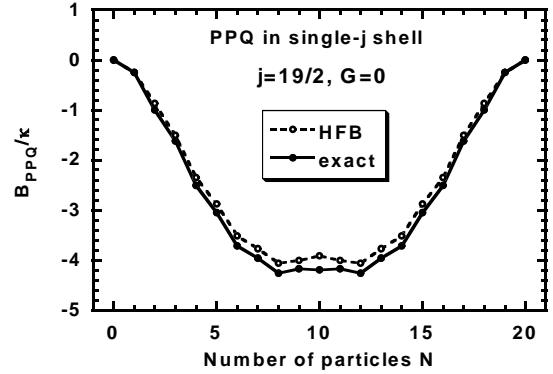

FIG. 7. Similar to Fig. 6 except for the energy-spacing filters (6) $\delta e_{\rm def}$. The single-particle spacings are marked by dots.

FIG. 9. Exact binding energies $B_{\rm PPQ}$ (solid line) of particles in the $j=19/2$ single-$j$ shell interacting with the pure QQ interaction ($G=0$). Energies obtained within the HFB approximation are shown with the dashed line.

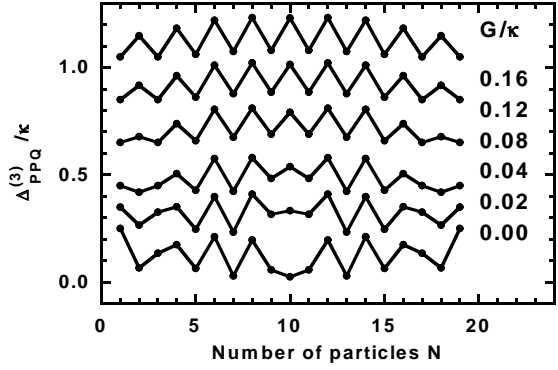

FIG. 10. Exact values of the three-mass filter $\Delta_{\text{PPQ}}^{(3)}$ within the $j=19/2$ PPQ calculated for the pairing strengths $G/\kappa$ indicated at the right-hand side.

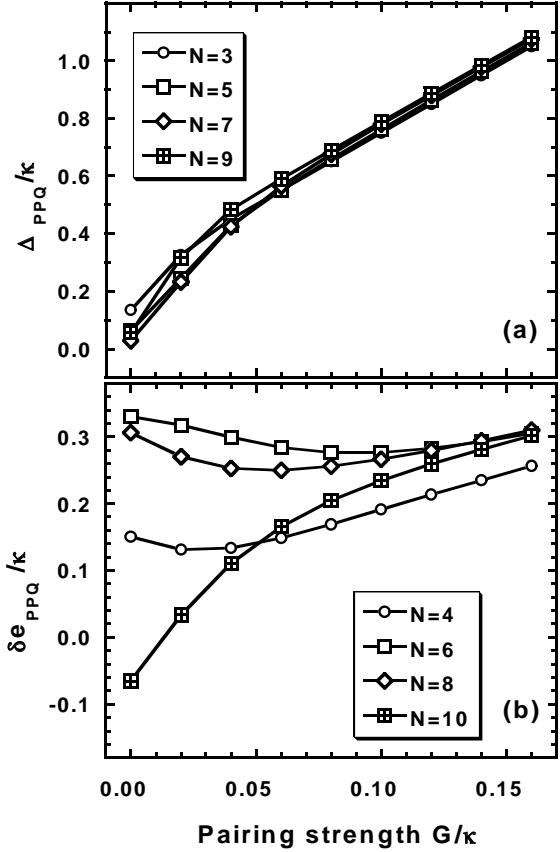

FIG. 11. Exact values of the OES $\Delta_{\text{PPQ}}$ (5) (a) and of the energy spacing $\delta e_{\text{PPQ}}$ (6b) (b) calculated within the $j=19/2$ PPQ model.

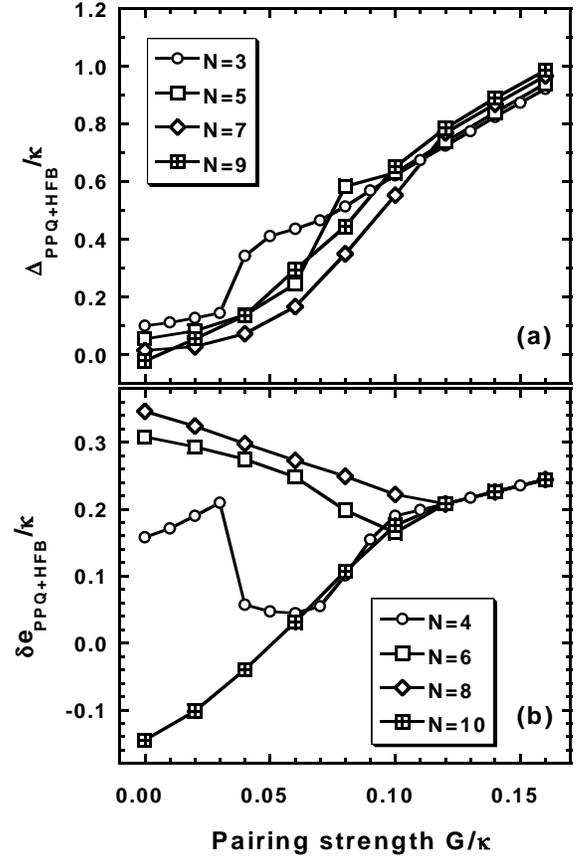

FIG. 12. Same as in Fig. 11 except for the HFB results.

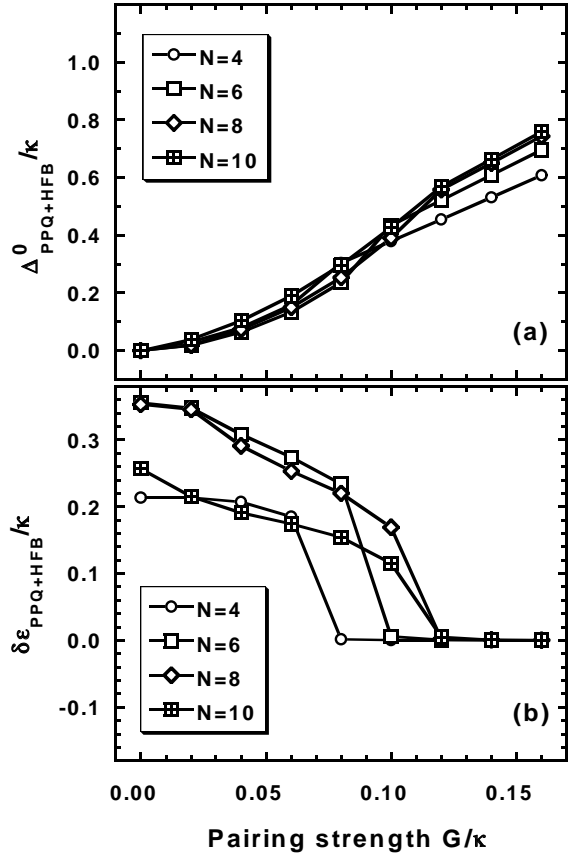

FIG. 13. Same as in Fig. 11 except for the $\Delta^0_{\text{PPQ+HFB}}$ HFB order parameter (45) and the differences of HFB canonical energies $\delta\varepsilon_{\text{PPQ+HFB}}$ (51).